\documentclass[doublespacing]{elsart}

\usepackage{graphicx}
\usepackage{amssymb}

\newcommand \beq  {\begin{equation}}
\newcommand \eeq  {\end{equation}}
\newcommand \bea {\begin{eqnarray} }
\newcommand \eea {\end{eqnarray}}
\newcommand \up{\uparrow}

\newcommand \rarrow{\rightarrow}
\newcommand \dg{^{\dagger}}
\newcommand \si { \sigma}
\newcommand \ra { \rangle}

\begin{document}
\begin{frontmatter}


\journal{SCES'2001: Version 1}


\title{Supersymmetric approach to the infinite U Hubbard Model}

%
%
%
%
%
%

\author[US]{P. Coleman}
\author[FR]{C.  P{\'e}pin}
%
 
\address[US]{Center for Materials Theory,
Department of Physics and Astronomy, 
Rutgers University, Piscataway, NJ 08854, USA.}
\address[FR]{ SPhT, L'Orme des Merisiers, CEA-Saclay, 91191 Gif-sur-Yvette, France.}

\begin{abstract}
We present a preliminary discussion on the 
use of supersymmetric representation of the Hubbard operator
which unifies the slave boson and slave fermion representations into a
single $U (1)\times SU (1\vert 1)$ gauge theory to treat the physics
of the infinite U Hubbard model. By looking for solutions
to the Hamiltonian in which the spins can exist as both
condensed ordered moments and as mobile charged carriers, we
examine the possibility of a Nagaoka ferromagnetic phase
at finite doping with a quantum critical point into the paramagnetic phase.

\end{abstract}

%
%



\end{frontmatter}

%
%
%
%
%

The infinite $U$ Hubbard model\cite{hubbard} is a prototype model for strongly
correlated electron systems. 
Interest in this venerable model has grown in recent times because
of its link with one of the basic models of high temperature
superconductivity, the $t-J$ model.\cite{tJreference} 
The infinite $U$ Hubbard model corresponds to the $t-J$ model without
antiferromagnetic interactions ($J\rarrow 0$). Despite the absence
of antiferromagnetic interactions, 
evidence from  finite temperature Lanzcos studies\cite{prevlosek} suggests that many of high temperature 
properties associated with the cuprate metals are already present in the
infinite $U$ Hubbard model.  

The $U=\infty$ Hubbard hamiltonian is written
\begin{equation} 
H_K= -\frac{t}{N} \sum_{(i,j)}[X_{\sigma 0} (i){\rm X}_{0\sigma }
(j)+{\rm H.c}]-\mu \sum_{j}{N}_{j} \ .  \end{equation} 
Here, the Hubbard
operators $X_{ab}= |a><b|$ 
where  $\vert a \rangle \in 
\{\vert 0\rangle, \{\vert \sigma\rangle\} \}$ describes 
a set of atomic  states involving a charged ``hole'' $\vert 0\rangle $  or a 
neutral spin state  $\vert \sigma \ra$ with  spin component
$\sigma \in \{1 \dots  N \}$ which for generality can have one of 
$N$ possible values.  $a, b \in {\vert 0\rangle , \vert \sigma \rangle }$

Thouless and Nagaoka~\cite{thouless,nagaoka} first established that the ground
state of the half filled $U= \infty$ Hubbard model doped with one hole
is a fully saturated ferromagnet. A wide body of theoretical
work\cite{kanamori,edwards,variational,putikka,sorella}
suggests that ferromagnetism remains stable to a finite hole doping, 
but there is no consensus about how it evolves and 
ultimately decays into  the paramagnetic state. In particular:
\begin{itemize}

\item Does a fully polarized state with  $M=1-x$ survive to a
finite doping, or is $M< (1-x)$ for all doping?

\item Is the transition to the paramagnet second  order with a quantum
critical point, or is it first order? 

\end{itemize}
For
the 2D square lattice 
variational studies 
give a critical doping $ \delta_c \simeq
0.22 \relbar 0.49$ \cite{variational}.
High temperature expansions~\cite{putikka} find that $\delta
_{c}\simeq 3/11$ with no region of fully saturated magnetization, 
yet recent variational Monte
Carlo studies~\cite{sorella} suggest that a partially polarized ferromagnet
survives between $\delta_1 \simeq 0.16$ and  a quantum critical point (QCP)
into the paramagnet at $\delta_2 \simeq 0.40 $.

There are two well-known approaches to incorporating the $U=\infty$ constraint
of the Hubbard operators:  the slave boson, and the slave fermion approach.\cite{tremblay} 
A mean-field slave fermion approach \cite{tremblay} 
leads to the conclusion that he infinite $U$ Hubbard model has a stable
ferromagnetic ground state for all doping, while using a slave bosons
representation, the system is paramagnetic for all dopings. 

In this
paper we use a supersymmetric representation of Hubbard
operators, given by
\begin{eqnarray}\label{therep}
X_{\sigma \sigma '} &=& b\dg _{\sigma }b_{\sigma '}+ f\dg _{\sigma
}f_{\sigma '}\cr
X_{\sigma 0}&=& b\dg _{\sigma }\chi + f\dg _{\sigma }\phi , \qquad 
X_{0\sigma }= \chi \dg b _{\sigma } + \phi\dg f _{\sigma }\cr
X_{00}&=& \chi\dg \chi +\phi \dg \phi \ .
\end{eqnarray} This
unifies the slave bosons and slave fermion approach
into a  single $U(1) \times SU(1 \vert 1)$ gauge theory~\cite{us}. 
Two constraints make the representation irreducible 
\begin{eqnarray}\label{theconstraints}
Q & = & n_{{ b}}
+ n_{{ \phi}} + n_{{f}} + n_{{ \chi}} \nonumber \\
Y & = & n_{{\phi}} + n_{{f}} - (n_{{ b}} + n_{{\chi}}) +
\frac{1}{Q}[\theta, 
\theta^{\dagger}] \ ,
\end{eqnarray} where
$\theta = \sum_{\sigma}{b}_{\sigma }^{\dagger}{f}_{\sigma } -
{\chi}^{\dagger}{\phi}$. The operators $\theta $ ,$\theta^\dagger$,
$Q$ and $Y$
commute
with the constraints and Hubbard operators, generating 
a local 
$SU (1|1)$ supersymmetry. 
For the  $S=1/2$ Hubbard model, we must choose $Q=1$ and $Y=0$.
A new feature is the
appearance of a ``superspin''
\[
\tau _{3} = \frac{1}{Q}[\theta\dg  ,\theta  ]
\]
which takes the values $+1$ and $-1$ in pure slave boson ($+1$) or slave
fermion  representation ($-1$), lying between
these extremes 
in a partially polarized ferromagnet. 

Consider the 
coherent state
$\vert b,\phi \rangle 
 = \exp [\sum_{j}b \hat b\dg _{j\up }+ \phi \hat
\phi \dg _{j}]\vert 0\rangle $  in which the  ``up'' Schwinger boson
and slave boson are condensed, 
where $\vert 0\rangle $ is the state
with one $\chi $ fermion per site and $b$ and $\phi $ are c-numbers. 
The action of the 
Hubbard operator on $\vert b,\phi \rangle $
is given by
\begin{eqnarray}\label{}
X_{\sigma o} ({\bf  k}
)
\vert b,\phi \rangle 
= (  \delta _{{\sigma \uparrow}}b\hat \chi \dg _{{\bf  k}}
 +
\phi \hat  f\dg _{{ {\bf  k}\sigma }})\vert b,\phi \rangle,
\end{eqnarray}
where $X_{\sigma o} (j)=\frac{1}{\sqrt{N}}\sum_{j}
X_{\sigma o} (j) e^{-i  {\bf k}\cdot {\bf x}_{j}}$ is the Fourier
transform of the Hubbard operator. 
If we now construct a Gutzwiller wavefunction with Fermi momenta
$k_{F\uparrow}$ and $k_{F\downarrow}$ for the up and down electrons,
we obtain
\begin{eqnarray}\label{gutz}
\vert \psi \rangle  &=& 
{ \mathcal P}_{G}\prod_{k<k_{F\sigma }, \sigma } X_{\sigma o} ({\bf  k})
\vert b,\phi \rangle\cr
&=&
{\mathcal P}_{G}\prod_{k<k_{F\sigma },\sigma } 
(b \chi \dg _{\vec{k }}\delta _{{\sigma \uparrow}} +
\phi f\dg _{\vec{ k \sigma }})\vert b,\phi \rangle 
\end{eqnarray}
where ${\mathcal P}_{G}$ is the projection operator that imposes $(Q,Y)= (1,0)$
on each site.

Here we outline an exploration of this wavefunction 
using analytic methods.  
To simplify matters 
we transform to a gauge
called the $\chi $ gauge  in which the
fermi field associated with the constraint is absorbed into the
slave fermion $\chi $. Details of this transformation will be provided
in a future publication.  The constraints in the $\chi $ gauge  for 
$S=1/2$ become
$Q=N_F + N_B = 1$ and $Y=N_F- N_B + \tau _{3} (\zeta) =   0$,
where $ N_F = n_f +n_{\varphi}$, $N_B= n_b + n_{\chi}$, $\tau _{3} (\zeta )= -\tanh(\zeta \beta) $ is the expectation
value of the superspin $ \hat \tau_3 $ and $\zeta$ is the chemical
potential conjugate to $Y$.
In other words \begin{eqnarray}
N_F & = &
\frac{1}{2} ( 1+ \tau _{3} (\zeta ))  \nonumber \\
 N_B  & = &  \frac{1}{2}( 1- \tau _{3} (\zeta )) \ .
\end{eqnarray} 
If $\zeta\ne 0$, then in the ground-state $\tau _{3}=-sgn (\zeta )$.
corresponding to a fully polarized slave fermion phase  or a
paramagnetic slave boson phase. 
However, the case $\lim_{T\rarrow 0} \zeta =0$ 
allows a new phase where 
both slave bosons and slave fermions coexist and 
$-1<\tau_3<1$.

Here we describe a  mean field approach
in which the 
constraint field $\chi $ is be approximated in a single sharp mode. We briefly describe the results of this analysis in 2D.
Carrying out a
particle-hole transformation on the field $\chi \rarrow  - \tilde
\chi\dg$, the mean field hamiltonian is
\begin{eqnarray}
H_{MF}& =&  \sum_{k} ( f \dg_{k, \si},\chi \dg _{\bf  k})
\left[\matrix{t_{\bf k}\phi ^{2}+\lambda_{f}& t_{\bf k}
\phi b\cr t_{\bf k}
&  t_k b^{2} + \lambda_\chi )}
 \right]
\left( \matrix{
f_{k, \si}  \cr \chi _{\bf k}}\right)\cr
& + & ( \lambda + \zeta ) \varphi^2 + ( \lambda -\zeta - \mu ) b^2 - \zeta (\tau_3 -1),
\end{eqnarray} where
$ t_k = -2 t \cos k_x + \cos k_y$, $\lambda_f= \lambda + \zeta - \mu$,
$\lambda_\chi = \zeta -\lambda $. Note 
the hybridization between the electron  
and spinless hole
$\chi$ which develops when $\phi b\ne 0$. 

We find that the mean-field Free energy $E-\mu N$ of
the paramagnetic 
($\zeta >0$) and fully polarized ground-state ($\zeta <0 $) become equal  at
a chemical potential $\mu^{*}= 0.77 t $, corresponding to two different
dopings $(\delta _{1},\delta _{2})= ( 0.25, 0.45)$. 
Varying $\tau_{3}$ at $\mu=\mu^{*}$, we
find that the point where $\zeta=dF/d\tau _{3}=0$ is a local maximum
(Fig 1 (b)),
corresponding to a \underline{phase separation}
between the paramagnetic and fully polarized phases between $\delta
_{1}$ and $\delta _{2}$. (Fig. 1 )
Unlike recent results of
Bocca and Sorella, there is no 
QCP at
$x=\delta _{2}$.  

Perhaps the most interesting result from this analysis, is the discovery
that the transition from the magnet to the paramagnet may occur via
a phase where $\zeta=0$, raising the fascinating possibility
that the magnetic quantum critical point involves soft supersymmetric
gauge fluctuations. 
In future work, we hope 
to examine whether the self-energy of the $\chi $ field and the
fluctuations in  $\zeta$ can stabilize a second-order QCP. 

We should particularly 
like to thank  A. M. Tremblay and J. Hopkinson for early discussions related to this
work. 
Discussions with W. Puttika and J. Hopkinson are also gratefully
acknowledged.  Part of this work was carried out at the Aspen Center
for Physics. 
This work was supported in part by the National Science Foundation
under grant DMR 9983156 (PC). 

%
%
%
%

%
%
%
%


 \begin{figure}[here]
     \centering
     \includegraphics{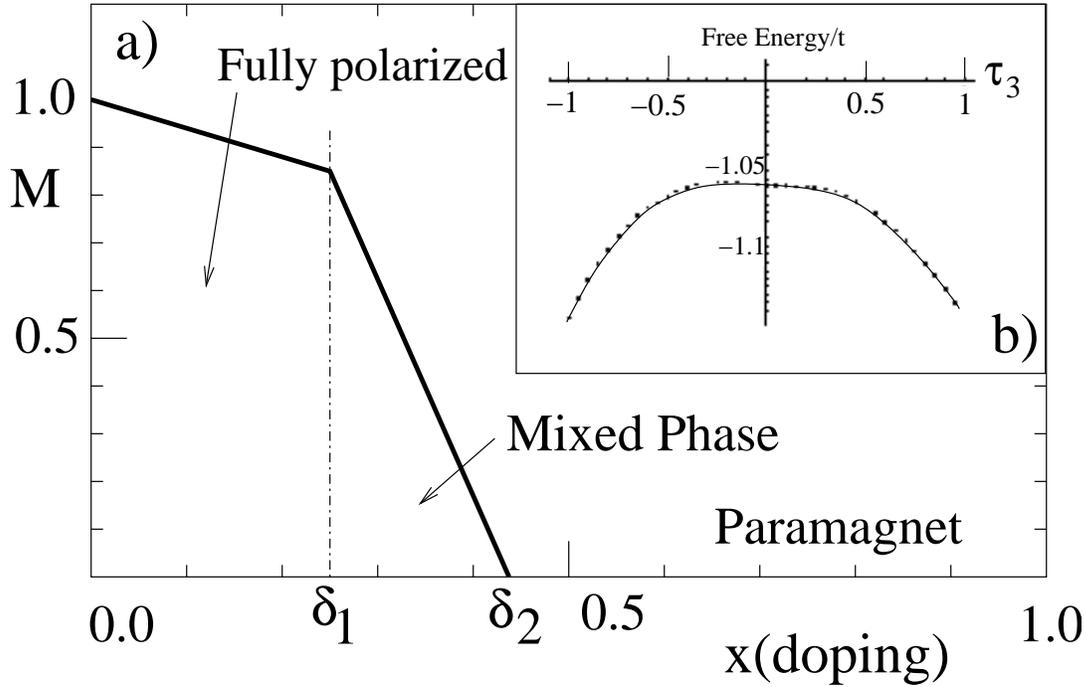}
     \caption{(a)Magnetization as a function of doping in mean field
theory. Inset (b) showing variation of free energy with $\tau _{3}$,
showing that a partially polarized state is unstable w.r.t.
phase separation. } 
 \end{figure}  
\end{document}